\documentclass[11pt]{amsart}

\usepackage{amsmath, amssymb, amsthm, amscd}
\usepackage{url}
\usepackage{graphicx}
\usepackage[hidelinks,pagebackref,pdftex]{hyperref}
\usepackage{color}
\usepackage{import}
\usepackage[margin=3cm]{geometry}

\usepackage{enumerate}
\usepackage[linesnumbered]{algorithm2e}
\usepackage{amsfonts}

 \usepackage{latexsym}

\newcommand{\tail}{last}

\newcommand{\rwit}{$\mbox{{\rm Wit}}^{\rho}$}
\newcommand{\wit}{{\rm Wit}}
\newcommand{\R}{\mathbb{R}}
\newcommand{\lk}{{\rm Lk}}
\newcommand{\K}{\mathcal{K}}
\newcommand{\Dim}[1]{\text{dim}(#1)}

\numberwithin{equation}{section}
\theoremstyle{plain}

\newtheorem{definition}[equation]{Definition}

\newtheorem{acknowledgements}[equation]{Acknowledgements}

\begin{document}

\title{The Simplex Tree: An Efficient Data Structure for General
  Simplicial Complexes}


\author{Jean-Daniel Boissonnat}
\author{Cl\'ement Maria}

\address{INRIA Sophia Antipolis-M\'editerran\'ee}
\email{jean-daniel.boisonnat@inria.fr, clement.maria@inria.fr}

\begin{abstract}

This paper introduces a new data structure, called simplex tree, to
  represent abstract simplicial complexes of any dimension. All faces
  of the simplicial complex are explicitly stored in a trie
  whose nodes are in bijection with the faces of the complex. This
  data structure allows to efficiently implement a large range of
  basic operations on simplicial complexes.
We provide theoretical complexity analysis as
  well as detailed experimental results. We more specifically study
Rips and witness complexes.

\bigskip

This article appeared in Algorithmica 2014~\cite{DBLP:journals/algorithmica/BoissonnatM14}. An extended abstract appeared in the proceedings of the European Symposium on Algorithms 2012~\cite{DBLP:conf/esa/BoissonnatM12}.


\end{abstract}

\maketitle

\section{Introduction}
\label{sec:intro}
Simplicial complexes are widely used in combinatorial and
computational topology, and have found many applications in topological data
analysis and geometric inference. A variety of simplicial
complexes have been defined, for example the \v{C}ech complex, the Rips
complex and the witness complex~\cite{Alexa_topologicalestimation,DBLP:books/daglib/0025666}.
However, the size of these structures grows
very rapidly with the dimension of the data set, and
their use in real applications  has been quite limited so far.

We are aware of only a few works on the design of data structures for
general simplicial complexes. Brisson~\cite{DBLP:conf/compgeom/Brisson89} and Lienhardt~\cite{DBLP:journals/ijcga/Lienhardt94}
have introduced data structures to represent $d$-dimensional cell
complexes, most notably subdivided manifolds. While those data
structures have nice algebraic properties, they are very redundant and
do not scale to large data sets or high dimensions.
Zomorodian~\cite{DBLP:conf/compgeom/Zomorodian10} has
proposed the tidy set, a compact data structure to simplify a
simplicial complex and compute its homology. Since the construction of
the tidy set requires to compute the maximal faces of the simplicial
complex, the method is especially designed 
for flag complexes. Flag complexes are a special type of simplicial
complexes (to be defined later) whose combinatorial structure can be
deduced from its graph. In particular, maximal faces of a flag complex
can be computed without
constructing explicitly the whole complex.
In the same spirit, Attali et al.~\cite{DBLP:conf/compgeom/AttaliLS11a}
have proposed the skeleton-blockers data structure. Again, the
representation is general but it requires to compute blockers, the
simplices which are not contained in the simplicial complex but whose
proper subfaces are. Computing the blockers is difficult in general and
details on the construction are given only for flag complexes, for
which blockers can be easily obtained.
As of now, there is no data structure for general
simplicial complexes that scales to dimension and size. The best
implementations have been restricted to flag complexes.

Our approach aims at combining both generality and scalability. We
propose a tree representation for simplicial complexes. The nodes of
the tree are in bijection with the simplices (of all dimensions) of
the simplicial complex. In this way, our data structure, called a
\emph{simplex tree}, explicitly stores all the
simplices of the complex but does not represent explicitly all the
adjacency relations between the simplices, two simplices being
adjacent if they share a common subface. Storing all the simplices
provides generality, and the tree structure of our representation
enables us to implement basic operations on simplicial complexes
efficiently, in particular to retrieve incidence relations, \emph{ie}
to retrieve the faces that contain a given simplex or are contained
in a given simplex.

The paper is organized as follows. In section~\ref{subsec:DS}, we
describe the simplex tree and, in section~\ref{subsec:op_on_st}, we
detail the elementary operations on the simplex tree such as adjacency
retrieval and maintainance of the data structure upon elementary
modifications of the complex. In section~\ref{sec:examples}, we
describe and analyze the construction of flag complexes, witness
complexes and relaxed witness complexes. An algorithm for inserting
new vertices in the witness complex is also described. Finally,
section~\ref{sec:experiments} presents a thorough experimental
analysis of the construction algorithms and compares our
implementation with the softwares {\sc JPlex} and {\sc
Dionysus}. Additional experiments are provided in
appendix~\ref{sec:additional_expe}. 


\subsection{Background}

\label{subsec:background}
\paragraph{Simplicial complexes.}
A {\em simplicial complex} is a pair $\K=(V,S)$ where $V$ is
a finite set whose elements are called the {\em vertices} of $\K$ and
$S$ is a set of non-empty subsets of $V$ that is
required to satisfy the following two conditions~:
\begin{enumerate}
\item $p\in V \Rightarrow \{  p\} \in S$
\item $\sigma\in S, \tau\subseteq \sigma \Rightarrow
\tau\in S$
\end{enumerate}
Each element $\sigma\in S$ is called a {\em simplex} or a \emph{face} of $\K$
and, if $\sigma\in S$ has precisely $s+1$ elements ($s\geq -1$),
$\sigma$ is called an $s$-simplex and the dimension of $\sigma$ is
$s$. The dimension of the simplicial complex $\K$ is
the largest $k$ such that $S$ contains a $k$-simplex.

We define the \emph{$j$-skeleton}, $j\geq 0$, of a simplicial complex $\K$ to be
the simplicial complex made of the faces of $\K$ of dimension
at most $j$. In particular, the $1$-skeleton of $\K$ contains
the vertices and the edges of $\K$. The $1$-skeleton has the
structure of a graph, and we will equivalently talk about the graph of
the simplicial complex.

A \emph{subcomplex} $\K' = (V',S')$ of the simplicial complex $\K =
(V,S)$ is a simplicial complex satisfying $V' \subseteq V$ and $S'
\subseteq S$. In particular, the $j$-skeleton of a simplicial complex is
a subcomplex.

\paragraph{Faces and cofaces.} 
A {\em face} of a simplex $\sigma=\{p_0 ,\cdots ,p_s\}$ is a simplex whose vertices
form a subset of $\{p_0 ,\cdots ,p_s\}$. A {\em proper} face
is a face different from $\sigma$ and the {\em facets} of $\sigma$ are
its proper faces of maximal dimension.
A simplex $\tau\in \K$ admitting $\sigma$ as a face is called a {\em
coface} of $\sigma$. The subset of simplices consisting of all the cofaces of a
simplex $\sigma \in \K$ is called the {\em star} of $\sigma$.

The {\em link}  of a simplex $\sigma$ in a simplicial complex $\K = (V,S)$ is
defined as the set of faces: 
$$\lk (\sigma) = \{ \tau\in S | \sigma\cup \tau \in S, \sigma \cap \tau = \emptyset \}$$ 

\paragraph{Filtration.}
A {\em filtration} over a simplicial complex $\K$ is an ordering of the
simplices of $\K$ such that all prefixes in the ordering are subcomplexes
of $\K$. In particular, for two simplices $\tau$ and $\sigma$ in the simplicial
complex such that $\tau \subsetneq \sigma$, $\tau$ appears before
$\sigma$ in the ordering. Such an ordering may be given by a real number associated to
the simplices of $\K$. The order of the simplices is simply the order
of the real numbers.

\section{Simplex Tree}
\label{sec:simplex-tree}

In this section, we introduce a new data structure which can represent
any simplicial complex. This data structure is a
trie~\cite{DBLP:conf/soda/BentleyS97} which explicitly represents all the simplices and allows
efficient implementation of basic operations on simplicial complexes.

\subsection{Simplicial Complex and Trie}
\label{subsec:DS}

Let  $\K = (V,S)$ be a simplicial complex of dimension $k$.
The vertices are labeled from $1$ to $|V|$ and ordered accordingly.

We can thus
associate to each simplex of $\K$ a word on the alphabet
$1 \cdots |V|$. Specifically, a $j$-simplex of $\K$ is uniquely
represented as the word of length $j+1$ consisting of the ordered set of the labels of
its $j+1$ vertices.  Formally, let simplex $\sigma = \{v_{\ell_0},
\cdots , v_{\ell_j}\} \in S$, where $v_{\ell_i}\in V$, $\ell_i\in \{1, \cdots ,|V|\}$ and
$\ell_0<\cdots < \ell_j$. $\sigma$ is then represented by the word
$[\sigma] = [\ell_0, \cdots , \ell_j]$. The last label of the word representation of a simplex $\sigma$ will
be called the last label of $\sigma$ and denoted by
$\tail (\sigma)$.

The simplicial complex $\K$ can be defined as a collection
of words on
an alphabet of size $|V|$. To compactly represent the set of simplices
of $\K$, we store the corresponding words in a tree satisfying
the following properties:

\begin{enumerate}
\item The nodes of the simplex tree are in bijection with the
  simplices (of all dimensions) of the complex. The
  root is associated to the empty face.
\item Each node of the tree, except the root, stores the label of a
  vertex. Specifically, a node associated to a simplex $\sigma \neq
  \emptyset$ stores the label $\tail (\sigma)$. 
\item The vertices whose labels are encountered along a path from the
  root to a node associated to a simplex $\sigma$, are the vertices of
  $\sigma$. 
 Along such a path, the labels are
  sorted by
  increasing order and each label appears no more than once.

\end{enumerate}

\begin{figure}
\begin{center}
\includegraphics[width=11cm]{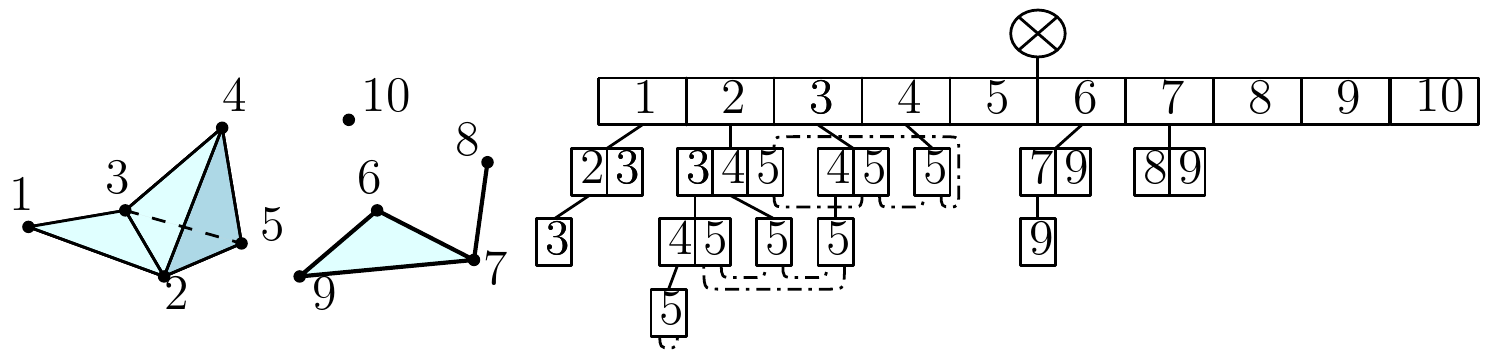}
\end{center}
\caption{A simplicial complex on $10$ vertices and its simplex
  tree. The deepest node represents the tetrahedron of the complex. All
  the positions of a given label at a given depth are linked in a
  list, as illustrated in the case of label $5$.}
\label{fig:preft}
\end{figure}

We call this data structure the \emph{Simplex Tree} of $\K$. It may be seen as a
trie~\cite{DBLP:conf/soda/BentleyS97} on the words
representing  the simplices of the complex (Figure~\ref{fig:preft}). The depth of the root is $0$ and the depth of a node
is equal to the dimension of the simplex it represents plus one.

In addition, we augment the data
structure so as to quickly locate all the instances of a given label
in the tree. Specifically,
all the nodes at a same depth $j$ which contain a same label $\ell$ are
linked in a circular list $L_j(\ell)$, as illustrated in Figure~\ref{fig:preft}
for label $\ell = 5$.

The children of the root of the simplex tree are called the
{\em top nodes}. The top nodes are in bijection with the elements of
$V$, the vertices of $\K$. Nodes which share the same
parent (e.g. the top nodes) will be called  \emph{sibling
  nodes}.

We also attach to each set of sibling nodes a pointer to their parent
so that we can access a parent in constant time.

We give a constructive definition of the simplex tree. Starting from
an empty tree, we insert the words
representing the simplices of the complex in the following
manner. When inserting the word $[\sigma]= [\ell_0,
\cdots , \ell_j]$ we start from the root, and follow the path containing
successively all labels $\ell_0, \cdots, \ell_i$, where $[\ell_0, \cdots , \ell_i]$
denotes the longest prefix of $[\sigma]$  already stored in the simplex tree. We then append
to the node representing $[\ell_0, \cdots , \ell_i]$ a path consisting of
the nodes storing
labels $\ell_{i+1}, \cdots , \ell_j$.

It is easy to see that the three properties above are
satisfied. Hence, if  $\K$ consists of $|\K|$
simplices
 (including the empty face), the associated
simplex tree contains exactly $|\K|$ nodes.

We use dictionaries with size linear in the
number of elements they store (like a red-black tree or a hash table)
for searching, inserting and removing elements
among a set of sibling nodes. Consequently these additional structures do not
change the asymptotic memory complexity of the simplex tree. For the
top nodes, we simply use an array since the set of vertices $V$ is
known and fixed. Let $\text{deg}(\mathcal{T})$ denote the maximal outdegree of a node,
in the simplex tree $\mathcal{T}$, distinct from the root. Remark that $\text{deg}(\mathcal{T})$ is at most the maximal
degree of a vertex in the graph of the simplicial complex. 
In the following, we will denote by $D_{\text{m}}$ the
maximal number of operations needed to perform a search, an insertion
or a removal in a dictionary of maximal size
$\text{deg}(\mathcal{T})$ (for example, with red-black trees $D_{\text{m}} =
O(\log(\text{deg}(\mathcal{T})))$ worst-case, with hash-tables $D_{\text{m}} =
O(1)$ amortized). Some algorithms, that we describe
later, require to intersect and to merge sets of
sibling nodes. In order to compute fast set operations, we will prefer
dictionaries which allow to traverse their elements in sorted order (e.g., red-black
trees). 
We discuss the value of $D_{\text{m}}$ at the end of this section in
the case where the points have a geometric structure.

We introduce two new notations for the analysis of the complexity of
the algorithms. Given a simplex $\sigma \in \K$, we define $C_\sigma$ to be
the number of cofaces of $\sigma$. Note that $C_\sigma$ only depends on the
combinatorial structure of the simplicial complex $\K$. Let
$\mathcal{T}$ be the simplex tree associated to $\K$. Given a label
$\ell$ and an index $j$, we define $\mathcal{T}_\ell^{> j}$ to be the number
of nodes of $\mathcal{T}$ at depth strictly greater than $j$ that store label
$\ell$. These nodes represent the simplices of dimension at least $j$ that
admit $\ell$ as their last label. $\mathcal{T}_\ell^{> j}$ depends on the labelling of the
vertices and is bounded by $C_{\{v_\ell\}}$, the number of cofaces of
the vertex with label $\ell$. For example, if $\ell$ is the greatest
label, we have $\mathcal{T}_{\ell}^{> 0} = C_{\{v_\ell\}}$, and if
$\ell$ is the smallest label we have $\mathcal{T}_{\ell}^{> 0} = 1$ independently from the number of cofaces of $\{v_\ell\}$.

\subsection{Operations on a Simplex Tree}
\label{subsec:op_on_st}

We provide algorithms for:

\begin{itemize}
\item {\sc Search/Insert/Remove-simplex} to search, insert or remove a
  single simplex, and {\sc Insert/Remove-full-simplex} to insert a
  simplex and its subfaces or remove a simplex and its cofaces
\item {\sc Locate-cofaces} to locate the cofaces of a simplex
\item {\sc Locate-facets} to locate the facets of a simplex
\item {\sc Elementary-collapse} to proceed to an elementary collapse
\item {\sc Edge-contraction} to proceed to contract an edge
\end{itemize}

\subsubsection{Insertions and Adjacency Retrieval}

\paragraph{Insertions and Removals}
Using the previous top-down traversal, we can \emph{search} and
\emph{insert} a word of length $j$ in $O(jD_{\text{m}})$ operations. 

We can extend this algorithm so as to insert a simplex and all its
subfaces in the simplex tree. Let $\sigma$ be a simplex we want to
insert with all its subfaces. Let $[\ell_{0}, \cdots ,\ell_j]$ be its
word representation. For $i$ from $0$ to $j$ we insert, if not already
present, a node $N_{\ell_i}$, storing label $\ell_i$, as a child of the
root. We recursively call the algorithm on the subtree rooted at $N_{\ell_i}$
for the insertion of the suffix $[\ell_{i+1}, \cdots ,\ell_j]$.
Since the number of subfaces of a simplex of dimension $j$ is
$\sum_{i=0\cdots j+1} \binom{j+1}{i} = 2^{j+1}$, this algorithm takes time
$O(2^jD_{\text{m}})$.

We can also remove a simplex from the simplex tree. Note that to
keep the property of being a simplicial complex, we need to remove all
its cofaces as well. We locate them thanks to the algorithm described below.

\paragraph{Locate cofaces.}
Computing the cofaces of a face is required to retrieve adjacency
relations between faces. In particular, it is useful when
traversing the complex or when removing a face. 
We also need to compute the cofaces of a face when contracting an edge (described later) or during the construction of
the witness complex, described later in section~\ref{subsec:witcpx}.

If $\tau$ is represented by the word $[\ell_0 \cdots \ell_j]$, the cofaces of $\tau$ are
the simplices of $\K$ which are represented by words of the form $[\star \ell_0
\star \ell_1 \star \cdots \star \ell_j \star]$, where $\star$
represents an arbitrary word on the alphabet, possibly empty.

To locate all the words of the form $[\star \ell_0 \star
\ell_1 \star \cdots \star \ell_j \star ]$ in the simplex tree, we first find all the words of the form $[\star \ell_0 \star \ell_1
\star \cdots \star \ell_j ]$. Using the lists $L_i(\ell_j)$ ($i >
j$), we find
all the nodes at depth at least $j+1$ which contain label $\ell_j$. For
each such node $N_{\ell_j}$, we traverse the tree upwards from $N_{\ell_j}$, looking
for a word of the form $[\star \ell_0 \star \ell_1 \star \cdots \star \ell_j
]$. If the search succeeds, the simplex represented by $N_{\ell_j}$ in the
simplex tree is a coface of $\tau$, as well as all the simplices
represented by the nodes in the subtree rooted at $N_{\ell_j}$, which
have word representation of the form $[\star \ell_0 \star \ell_1 \star \cdots \star \ell_j
\star]$. Remark that the cofaces of a simplex are represented by
a set of subtrees in the simplex tree. The procedure searches only
the roots of these subtrees.

The complexity for searching the cofaces of a simplex $\sigma$ of
dimension $j$ depends on the
number $\mathcal{T}_{\tail(\sigma)}^{>j}$ of nodes with label $\tail{ (\sigma) }$ and depth at
least $j+1$. If $k$ is the dimension of the simplicial complex,
traversing the tree upwards takes $O(k)$ time.
The complexity of this procedure is thus $O(k\mathcal{T}_{\tail(\sigma)}^{>j})$.

\paragraph{Locate Facets.}

Locating the facets of a simplex efficiently is the key point of
the incremental algorithm we use to construct witness complexes in
section~\ref{subsec:witcpx}. 

\begin{figure}
\begin{center}
\includegraphics[width=8cm]{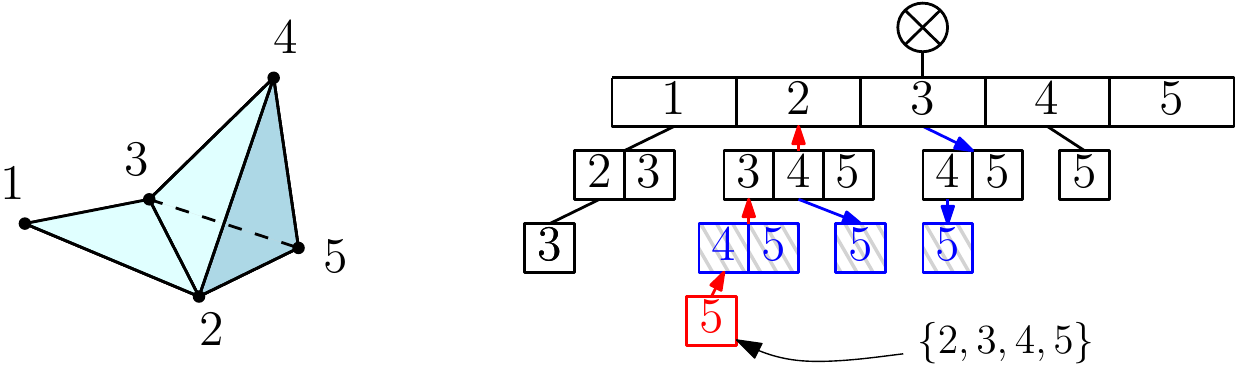}
\end{center}
\caption{Facets location of the simplex $\sigma = \{2,3,4,5\}$,
  starting from the position of $\sigma$ in the simplex tree. The
  nodes representing the facets are colored in grey.}
\label{fig:algo_facets}
\end{figure}

Given a simplex $\sigma$, we want to access the nodes of the simplex
tree representing the facets of $\sigma$.  If the word
representation of $\sigma$ is $[\ell_0,\cdots,\ell_{j}]$, the word
representations of the facets of $\sigma$ are 
the words $[\ell_0, \cdots , \widehat{\ell}_i, \cdots , \ell_j]$, $0 \leq i
\leq j$, where $\widehat{\ell}_i$ indicates that $\ell_i$ is omitted. 
If we denote, as before, $N_{\ell_i}, i=0,\cdots ,j$ the nodes representing the words $[\ell_0, \cdots ,
\ell_{i}],i=0,\cdots ,j$ respectively, a traversal from the node representing $\sigma$ up to the
root will exactly pass through the nodes $N_{\ell_i}$, $i = j, \cdots ,0$. When
reaching the node $N_{\ell_{i-1}}$, a search from $N_{\ell_{i-1}}$ downwards for the word $[\ell_{i+1}, \cdots ,
\ell_{j}]$ locates (or proves the absence of) the facet $[\ell_0, \cdots
, \widehat{\ell}_i, \cdots , \ell_j]$. See Figure~\ref{fig:algo_facets} for a running example.

This procedure locates all the facets of the $j$-simplex $\sigma$ in $O(j^2
D_{\text{m}})$ operations. 

\paragraph{Experiments.}

We report on the experimental performance of the facets and cofaces
location algorithms. Figure~\ref{fig:time_fac_cof}
represents the average time for these operations on a simplex, as a
function of the dimension of the simplex. We use the dataset {\bf
  Bro}, consisting of points in $\mathbb{R}^{25}$, on top of which we
build a relaxed witness complex with $300$ landmarks and $15,000$
witnesses, and relaxation parameter $\rho = 0.15$. See
section~\ref{sec:experiments} for a detailed description of the
experimental setup. We obtain a $13$-dimensional
simplicial complex with $140,000$ faces in less than $3$ seconds. 

\setlength{\tabcolsep}{2.5pt}

\begin{figure}
\begin{tabular}{|c|r|r|r|r|r|r|r|r|r|r|r|r|r|r|}
 \hline
 Dim.Face&0&1&2&3&4&5&6&7&8&9&10&11&12&13\\ 
 \hline
$\#$ Faces&300&2700&8057&15906&25271&30180&26568&17618&8900&3445&1015&217&30&2\\
 \hline
\end{tabular}

\setlength{\tabcolsep}{9pt}
 \begin{tabular}{c c}
    \begin{minipage}[b]{0.45\linewidth}
      \centering
      \includegraphics{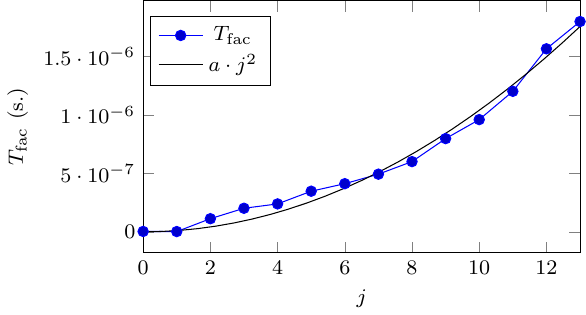}
    \end{minipage}
    &  
    \begin{minipage}[b]{0.45\linewidth}
      \centering
      \includegraphics{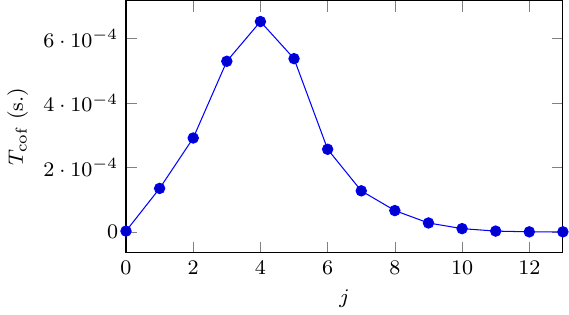}
    \end{minipage}\\
  \end{tabular}

\caption{Repartition of the number of faces per dimension (top) and average time to compute the facets (left) and the
  cofaces (right) of a simplex of a given dimension.} 
\label{fig:time_fac_cof}
\end{figure}

The theoretical
complexity for computing the facets of a $j$-simplex $\sigma$ is
$O(j^2 D_{\text{m}})$. As reported in Figure~\ref{fig:time_fac_cof}, the
average time to search all facets of a $j$-simplex is well
approximated by a quadratic function of the dimension $j$ (the standard
error in the approximation is $2.0\%$).

A bound on the complexity of computing the cofaces of a $j$-simplex
$\sigma$ is $O(k\mathcal{T}_{\tail(\sigma)}^{>j})$, where
$\mathcal{T}_{\tail(\sigma)}^{>j}$ stands for the number of nodes in
the simplex tree that store the label $\tail{(\sigma)}$ and have depth
larger than $j+1$. Figure~\ref{fig:time_fac_cof} provides experimental results
for a random labelling of the vertices.  As can be seen, the time for
computing the cofaces of a simplex $\sigma$ is low, on average, when
the dimension of $\sigma$ is either small ($0$ to $2$) or big ($6$ to
$13$), and  higher for  intermediate dimensions ($3$ to $5$). The
value $\mathcal{T}_{\tail(\sigma)}^{>j}$ in the complexity analysis
depends on both the labelling of the vertices and the number of cofaces of the
vertex $v_{\tail{(\sigma)}}$: these dependencies make the analysis of
the algorithm quite difficult, and we let as an open problem to fully understand
the experimental behavior of the algorithm as observed in
Figure~\ref{fig:time_fac_cof}
(right).


\subsubsection{Topology preserving operations}

We show how to implement two topology preserving operations on a
simplicial complex represented as a simplex tree. Such simplifications
are, in particular, important in topological data analysis.

\paragraph{Elementary collapse.} 

We say that a simplex $\sigma$ is collapsible through one of
its faces $\tau$ if $\sigma$ is the only coface of $\tau$, which can
be checked by computing the cofaces of $\tau$. Such a pair $(\tau,\sigma)$
is called a \emph{free pair}. Removing both faces of a free pair is an
elementary collapse.

Since $\tau$ has no coface other than $\sigma$, either the node representing
$\tau$ in the simplex tree is a leaf (and so is the node
representing $\sigma$), or it has the node
representing $\sigma$ as its unique child. An elementary collapse of
the free pair $(\tau,\sigma)$ consists either in the removal of the
two leaves representing $\tau$ and $\sigma$, or the removal of the
subtree containing exactly two nodes: the node representing $\tau$ and
the node representing $\sigma$.

\paragraph{Edge contraction.}
Edge contractions are used in~\cite{DBLP:conf/compgeom/AttaliLS11a} as
a tool for homotopy preserving simplification and
in~\cite{DBLP:journals/corr/abs-1208-5018} for computing the
persistent topology of data points. 
Let $\K$ be a simplicial complex and let $\{v_{\ell_a},v_{\ell_b}\}$ be an edge
of $\K$ we want to contract. We say that we contract $v_{\ell_b}$ to $v_{\ell_a}$ meaning that $v_{\ell_b}$ is
removed from the complex and the link of $v_{\ell_a}$ is augmented with
the link of $v_{\ell_b}$. Formally, we define the map $f$ on the set of vertices $V$ which
maps $v_{\ell_b}$ to $v_{\ell_a}$
and acts as the identity function for all other inputs:

\[
  f(u) = \left\{
          \begin{array}{ll}
            v_{\ell_a} & \qquad \mathrm{if}\quad u = v_{\ell_b} \\
            u & \qquad \mathrm{otherwise} \\
          \end{array}
        \right.
\]

We then extend $f$ to all simplices $\sigma = \{v_{\ell_0}, \cdots , v_{\ell_j}\}$
of $\K$ with $f(\sigma) = \{f(v_{\ell_0}), \cdots , f(v_{\ell_j})\}$. The
contraction of $v_{\ell_b}$ to $v_{\ell_a}$ is defined as the operation which replaces
$\K = (V,S)$ by $\K' = (V \setminus \{v_{\ell_b}\} , \{f(\sigma) | \sigma \in S)
\}$. $\K'$ is a simplicial complex. 

It has been proved in~\cite{DBLP:conf/compgeom/AttaliLS11a} that
contracting an edge $\{v_{\ell_a},v_{\ell_b}\}$
preserves the homotopy type of a simplicial complex whenever the
\emph{link condition} is satisfied:
\[
\lk (\{v_{\ell_a},v_{\ell_b}\}) = \lk( \{v_{\ell_a}\}) \cap
\lk(\{v_{\ell_b}\})
\]
This link condition can be checked using the {\sc Locate-cofaces}
algorithm described above.

Let $\sigma$ be a
simplex of $\K$. We
distinguish three cases~: 1. $\sigma$ does not
contain $v_{\ell_b}$ and remains unchanged; 2. $\sigma$ contains both $v_{\ell_a}$ and
$v_{\ell_b}$, and $f(\sigma) = \sigma \setminus \{v_{\ell_b}\}$; $|f(\sigma)| = |\sigma|
-1$ and  $f(\sigma)  $ is a strict subface of
$\sigma$; 3. $\sigma$ contains $v_{\ell_b}$ but not $v_{\ell_a}$ and $f(\sigma) = \left(\sigma
\setminus \{v_{\ell_b}\}\right) \cup \{v_{\ell_a}\}$, ($|f(\sigma)| = |\sigma|$).

\begin{figure}
\begin{center}
\includegraphics[width=11cm]{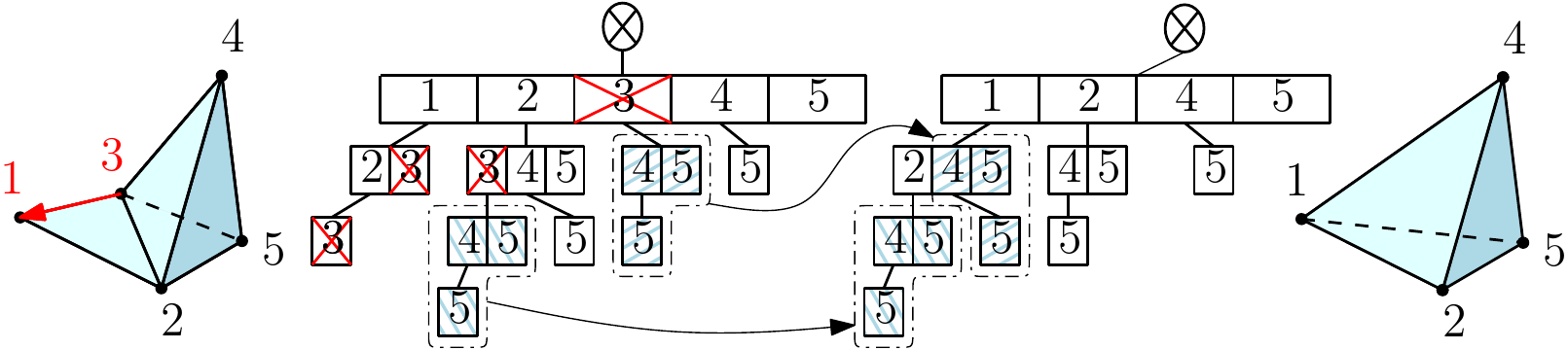}
\end{center}
\caption{Contraction of vertex $3$ to vertex $1$ and the associated modifications of
  the simplicial complex and of the simplex tree. The nodes which are
  removed are marked with a red cross, the subtrees which are moved
  are colored in blue.}
\label{fig:merge}
\end{figure}

We describe now how 
to compute the contraction of $v_{\ell_b}$ to $v_{\ell_a}$ when $\K$ is represented as
a simplex tree. We suppose that the edge $\{v_{\ell_a},v_{\ell_b}\}$ is in the complex
and, without loss of generality, $\ell_a < \ell_b$. All the simplices which do not contain
$v_{\ell_b}$ remain unchanged and we do not consider them. If a simplex $\sigma$ contains
both $v_{\ell_a}$ and $v_{\ell_b}$, it will become $\sigma \setminus
\{v_{\ell_b}\}$, after edge contraction, which
is a simplex already in $\K$. 
We simply remove $\sigma$ from the simplex
tree. Finally, if $\sigma$ contains $v_{\ell_b}$ but not $v_{\ell_a}$, we need to remove $\sigma$
from the simplex tree and add the new simplex $\left(\sigma
\setminus \{v_{\ell_b}\}\right) \cup \{v_{\ell_a}\}$.

We consider each node $N_{\ell_b}$ with label $\ell_b$ in turn. To do
so, we use the lists $L_j(\ell)$ which link all nodes
cointaining the label $\ell$ at depth $j$. Let
$\sigma$ be the simplex represented by $N_{\ell_b}$.
The algorithm traverses the tree upwards from $N_{\ell_b}$ and collects the
vertices of $\sigma$.  Let $T_{N_{\ell_b}}$
be the subtree rooted at $N_{\ell_b}$. As $\ell_a<\ell_b$, if $\sigma$
contains both $v_{\ell_a}$ and $v_{\ell_b}$, this will be true for all the simplices
whose representative nodes are in $T_{N_{\ell_b}}$, and, if $\sigma$
contains only $v_{\ell_b}$, the same will be true for all the simplices
whose representative nodes are in $T_{N_{\ell_b}}$. Consequently, if
$\sigma$ contains both $v_{\ell_a}$ and $v_{\ell_b}$, we remove the whole subtree $T_{N_{\ell_b}}$
from the simplex tree. Otherwise, $\sigma$ contains only $v_{\ell_b}$, all words
represented in $T_{N_{\ell_b}}$ are of the form $[\sigma '] \centerdot [\sigma''] \centerdot [\ell_b]
\centerdot [\sigma''']$ and will be turned into words $[\sigma'] \centerdot [\ell_a] \centerdot
[\sigma''] \centerdot [\sigma''']$ after edge contraction. We then have to move the subtree $T_{N_{\ell_b}}$
(except its root) from position $[\sigma '] \centerdot [\sigma'']$ to position
$[\sigma ']\centerdot [\ell_a] \centerdot [\sigma'']$ in the simplex tree. If a subtree is
already rooted at this position, we have to merge $T_{N_{\ell_b}}$ with this subtree as illustrated in
Figure~\ref{fig:merge}. In order to merge the subtree $T_{N_{\ell_b}}$
with the subtree rooted at the node representing the word $[\sigma
']\centerdot [\ell_a] \centerdot [\sigma'']$, we can successively
insert every node of $T_{N_{\ell_b}}$ in the corresponding set of
sibling nodes, stored in a dictionary. See Figure~\ref{fig:merge}.

We analyze the complexity of contracting an edge
$\{v_{\ell_a},v_{\ell_b}\}$. For each node storing the label $\ell_b$, we
traverse the tree upwards. This takes $O(k)$ time if the
simplicial complex has dimension $k$. As there are
$\mathcal{T}_{\ell_b}^{>0}$ such nodes, the total cost is
$O(k \mathcal{T}_{\ell_b}^{>0})$.
We also manipulate the subtrees rooted at the nodes storing label
$\ell_b$. Specifically,  either we remove such a subtree or we move a
subtree by changing its parent node.  In the latter case, we have
to merge two subtrees. This is the more costly operation which takes,
in the worst case, $O(D_{\text{m}})$ operations per node in the
subtrees to be merged. As any node in such a subtree represents a coface of
vertex $v_{\ell_b}$, the total number of nodes in all the subtrees we
have to manipulate is at most $C_{\{v_{\ell_b}\}}$, and  the
manipulation of the subtrees takes $O(C_{\{v_{\ell_b}\}}
D_{\text{m}})$ time.
Consequently, the time needed to contract the edge
$\{v_{\ell_a},v_{\ell_b}\}$ is $O(k \mathcal{T}_{\ell_b}^{>0} + C_{\{v_{\ell_b}\}}
D_{\text{m}})$.

\paragraph{Remark on the value of $D_{\text{m}}$}: $D_{\text{m}}$
appears as a key value in the complexity analysis of the algorithms. Recall that $D_{\text{m}}$ is
the
maximal number of operations needed to perform a search, an insertion
or a removal in a dictionary of maximal size
$\text{deg}(\mathcal{T})$ in the simplex
tree. We suppose in the following that the dictionaries used are
red-black trees, in which case $D_{\text{m}} = O(\log (\text{deg}(\mathcal{T})))$. As mentioned earlier, $\text{deg}(\mathcal{T})$ is bounded by the maximal
degree of a vertex in the graph of the simplicial complex. In the worst-case,
if $n$ denotes the number of vertices of the simplicial complex, 
we have
$\text{deg}(\mathcal{T}) = O(n)$, and $D_{\text{m}} = O(\log(n))$.
However, this bound can be improved in the case of simplicial
complexes constructed on sparse data points
sampled from a low dimensional manifold, an important case in
practical applications.  Let $\mathbb{M}$ be a $d$-manifold with bounded
curvature, embedded in $\R^D$ and assume that the length of the longest (resp., shortest) edge of the
simplicial complex has length at most $r$ (resp., at least
$\epsilon$).
Then, a volume argument shows that the maximal degree of a vertex in the simplicial complex
is $\Theta((r/\epsilon)^d)$. 
Hence, when $r=O(\epsilon)$, which is a typical situation when $S$ is an $\epsilon$-net
of $\mathbb{M}$, the
value of $D_{\text{m}}$ is $O(d)$ with a constant depending only on
local geometric quantities.

\section{Construction of Simplicial Complexes}
\label{sec:examples}

In this section, we detail how to construct two important types of
simplicial complexes, the flag and the witness complexes, using simplex trees.

\subsection{Flag complexes}
 
A flag complex is a simplicial complex whose combinatorial structure is entirely
determined by its $1$-skeleton. Specifically, a simplex is in the flag
complex if and only if its
vertices form a clique in the graph of the simplicial complex, or,
in other terms, if and only if its
vertices are pairwise linked by an edge.

\paragraph{Expansion.} 

Given the $1$-skeleton of a flag complex, we call \emph{expansion of
  order $k$} the operation which reconstructs the $k$-skeleton of the flag complex. If the
$1$-skeleton is stored in a simplex tree, the expansion of order $k$
consists in successively inserting all the
simplices of the $k$-skeleton into the simplex tree.

Let $G = (V,E)$ be the graph of the simplicial complex, where
$V$ is the set of vertices and $E \subseteq V\times V$ is the set of
edges. For a vertex $v_{\ell} \in V$, we denote by
\[
\mathcal{N}^+(v_{\ell}) = \{\ell' \in \{1, \cdots ,|V|\} \mbox{ } | \mbox{ }
 (v_{\ell},v_{\ell'}) \in E \wedge \ell' > \ell\}
\]
the set of labels of the neighbors of
$v_{\ell}$ in $G$ that are bigger than $\ell$. 
Let $N_{\ell_j}$ be the node in the tree  that stores the label $\ell_j$ and
represents the word $[\ell_0, \cdots , \ell_j]$.  The
 children of $N_{\ell_j}$ store the labels in $\mathcal{N}^+(v_{\ell_0}) \cap \cdots \cap
\mathcal{N}^+(v_{\ell_j})$. Indeed, the children
of $N_{\ell_j}$ are  neighbors in $G$ of the vertices $v_{\ell_i}$, $0 \leq
i \leq j$, (by definition of a clique) and must have a bigger label
than $\ell_0, \cdots , \ell_j$ (by construction of the simplex tree).

\begin{figure}
\begin{center}
\includegraphics[width=9cm]{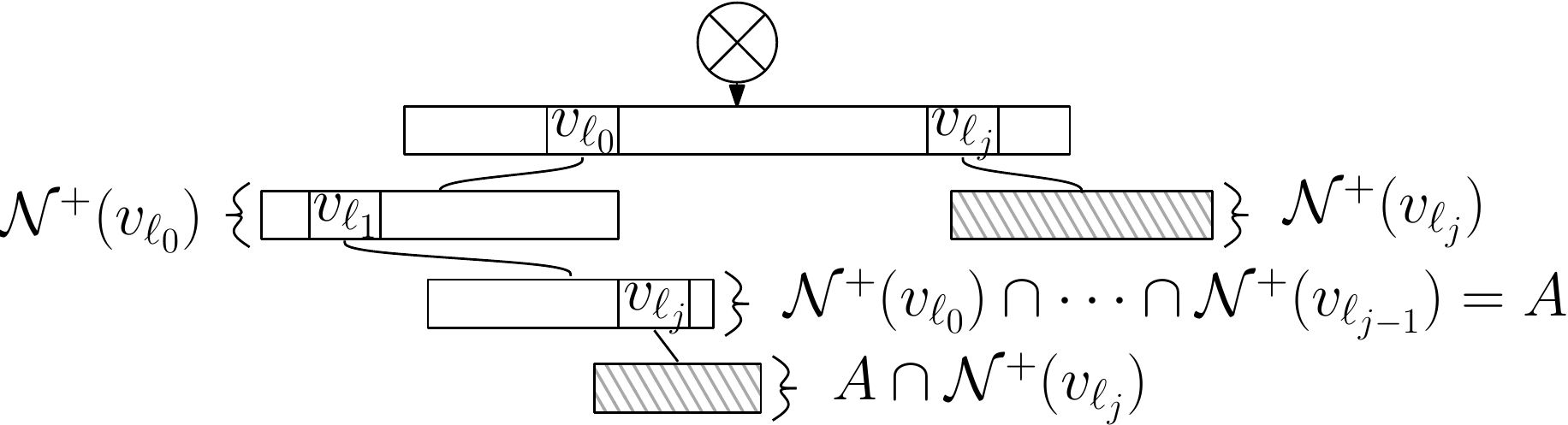}
\end{center}
\caption{Representation of a set of sibling nodes as intersection of neighborhoods.}
\label{fig:inter}
\end{figure}

Consequently, the sibling nodes of
$N_{\ell_j}$ are exactly the nodes that store the labels  in
$A=\mathcal{N}^+(v_{\ell_0})~\cap~\cdots~\cap~\mathcal{N}^+(v_{\ell_{j-1}})$,
and the children of $N_{\ell_j}$ are exactly the nodes that store the labels in $A\cap
\mathcal{N}^+(v_{\ell_j})$. See Figure~\ref{fig:inter}.

For every vertex $v_{\ell}$, we have an easy access to
$\mathcal{N}^+(v_{\ell})$ since $\mathcal{N}^+(v_{\ell})$ is exactly
the set of labels stored in the children of the top node storing label
$\ell$. We easily deduce an in-depth expansion algorithm.

The time complexity for the expansion algorithm depends on our 
ability to fastly compute intersections of the type $A \cap
\mathcal{N}^+(v_{\ell_j})$. In all of our experiments on the Rips
complex (defined below) we have observed that the time taken by the expansion algorithm
depends linearly on the size of the output simplicial complex, for a
fixed dimension. More details can be found in
section~\ref{sec:experiments} and appendix~\ref{sec:additional_expe}.

\paragraph{Rips Complex.} Rips complexes are geometric flag complexes
which are popular in computational topology due to their simple
construction and their good approximation
properties~\cite{DBLP:conf/compgeom/AttaliLS11,DBLP:conf/compgeom/ChazalO08}. Given a set of vertices $V$ in a
metric space and a parameter $r>0$, the Rips graph is defined as the
graph whose set of vertices is $V$ and two vertices are joined by an
edge if their distance is at most $r$. The Rips complex
is the flag complex defined on top of this graph. We will use this
complex for our experiments on the construction of flag complexes.

\subsection{Witness complexes}
\label{subsec:witcpx}

\paragraph{The Witness Complex.}
has been first introduced in~\cite{Alexa_topologicalestimation}.
Its definition involves two given sets of points in a metric space, the set of landmarks
$L$ and the set of witnesses $W$.

\begin{definition}
A witness $w \in W$ \emph{witnesses} a simplex $\sigma \subseteq L$ iff:
$$\forall x \in \sigma \mbox{ and } \forall y \in L\setminus \sigma
\mbox{ we have } \text{d}(w,x) \leq \text{d}(w,y)$$
\end{definition}

For simplicity of exposition, we will suppose that no
landmarks are at the exact same distance to a witness. In this case, a
witness $w \in W$ \emph{witnesses} a simplex $\sigma \subseteq L$ iff
the vertices of $\sigma$ are the $|\sigma|$ nearest neighbors of $w$ in $L$.
We study later
the construction of the \emph{relaxed witness complex}, which is a
generalization of the witness complex which includes the case where
points are not in general position.

The \emph{witness complex} $\wit(W,L)$ is the maximal
simplicial complex, with vertices in $L$, whose faces admit a witness
in $W$. Equivalently, a simplex belongs to the witness complex if and
only if it
is witnessed and  all its facets belong to  the witness
complex. A simplex satisfying this property will be called \emph{fully witnessed}. 

\paragraph{Construction Algorithm.}
We suppose the sets $L$ and $W$ to be finite and give them labels $\{1,
\cdots , |L|\}$ and $\{1, \cdots , |W|\}$ respectively. We describe
how to construct
the $k$-skeleton of the witness complex,  where $k$ may be any integer
in $\{1,\cdots , |L|-1\}$.

Our construction algorithm is incremental, from lower to higher
dimensions. At step $j$ we insert in the simplex tree the
$j$-dimensional fully witnessed simplices.

During the construction of the $k$-skeleton of the witness complex, we
need to access the nearest neighbors of the witnesses, in $L$. To do so, we compute the $k+1$ nearest neighbors
of all the witnesses in a preprocessing phase, and store them in a
$|W|\times (k+1)$ matrix. Given an index $j \in \{0, \cdots ,k\}$ and a witness
$w \in W$, we
can then access in constant time the $(j+1)^{\text{th}}$ nearest neighbor of $w$. 
We denote this landmark by $s_j^w$. We maintain a list of \emph{active witnesses}, initialized with
$W$. 
We insert the vertices of $\wit(W,L)$ in the simplex tree. For each
witness $w\in W$ we insert a top node storing the label of the nearest
neighbor of $w$ in $L$, if no such node already exists. $w$ is initially an
active witness and we make it point to the node mentionned above, representing the
$0$-dimensional simplex $w$ witnesses.

We maintain the following loop invariants: 
\begin{enumerate}
\item at the beginning of iteration $j$, the simplex tree
contains the $(j-1)$-skeleton of the witness complex
$\wit(W,L)$
\item the active witnesses are the elements of $W$ that
witness a
$(j-1)$-simplex of the complex; each active witness $w$ points to the node
representing the $(j-1)$-simplex in the tree it witnesses.
\end{enumerate}

\begin{figure}
\begin{center}
\includegraphics[width=11cm]{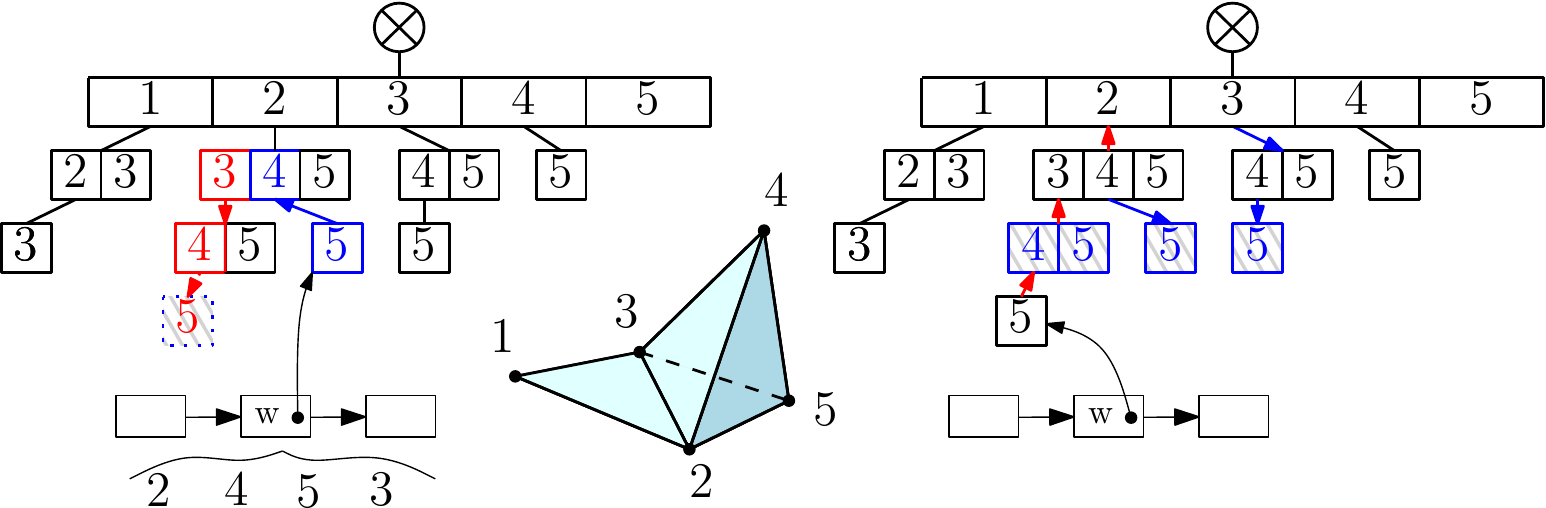}
\end{center}
\caption{Third iteration of the witness complex construction. The
  active witness $w$ witnesses the tetrahedron $\{2,3,4,5\}$ and points
to the triangle $\{2,4,5\}$. (Left) Search for the potential position
of the simplex $\{2,3,4,5\}$ in the simplex tree. (Right) Facets
location for simplex $\{2,3,4,5\}$, and update of the pointer of the
active witness $w$.}
\label{fig:algos_wc}
\end{figure}

At iteration $j\geq 1$, we traverse the list of active witnesses. Let
$w$ be an active witness. We first retrieve the $(j+1)^{\text{th}}$ nearest neighbor
$s_j^w$ of $w$ from the nearest neighbors matrix (Step 1).  Let $\sigma_j$ be the $j$-simplex 
witnessed by $w$ and let us decompose the word representing $\sigma_j$ into $[\sigma_j] = [\sigma']\centerdot
[s_j^w]\centerdot[\sigma'']$ (``$\centerdot$'' denotes the concatenation of words).
We then look for the location in
the tree where $\sigma_j$ might be inserted (Step 2).
To do so, we start at the node $N_w$ which represents the
$(j-1)$-simplex  witnessed by $w$. Observe
that the word associated to the path from the root to $N_w$ is exactly $[\sigma']\centerdot
[\sigma'']$. We walk $|[\sigma'']|$ steps up 
from $N_w$, reach the node representing $[\sigma']$ and then search
downwards for the word
$[s_w^j]\centerdot[\sigma'']$ (see Figure~\ref{fig:algos_wc}, left). The cost of this operation is
$O(jD_{\text{m}})$.

If the node representing $\sigma_j$ exists, $\sigma_j$ has already been inserted; we
update the pointer of $w$ and return. If the simplex tree contains neither this node
nor its father, $\sigma_j$ is not fully witnessed because the facet
represented by its longest prefix is missing. We consequently remove $w$
from the set of active witnesses.
Lastly, if the node is not in the tree but its father is, we check whether
$\sigma_j$ is fully witnessed.  To do so, we
search for the $j+1$ facets of $\sigma_j$ in the simplex tree (Step
3). The cost of this 
operation is $O(j^2D_{\text{m}})$ using the {\sc Locate-facets}
algorithm described in section~\ref{subsec:op_on_st}. If $\sigma_j$ is fully witnessed, we
insert $\sigma_j$  in the simplex tree and update the pointer of the active
witness $w$. Else, we remove $w$ from the list of active witnesses (see
Figure~\ref{fig:algos_wc}, right).

It is easily seen that the loop invariants are satisfied at the end of
iteration~$j$.

\paragraph{Complexity.}
The cost of accessing a neighbor of a witness using the nearest
neighbors matrix is $O(1)$.
We access a neighbor (Step 1) and locate a node in the simplex tree
(Step 2) at most $k|W|$ times. In total, the cost of Steps 1 and 2
together is
$O(|W|k^2 D_{\text{m}})$. In Step 3, either we insert
a new node in the simplex tree, which happens exactly $|\mathcal{K}|$ times
(the number of faces in the complex), or we remove an active
witness, which happens at most $|W|$ times. The total cost of Step
3 is thus $O( (|\mathcal{K}| + |W|)k^2D_{\text{m}})$. In conclusion, 
constructing the $k$-skeleton of the witness complex takes time
$$O((|\mathcal{K}| + |W|)k^2D_{\text{m}} + k|W|) = O((|\mathcal{K}| + |W|)k^2D_{\text{m}}).$$

\paragraph{Landmark Insertion.}

We present an algorithm to update the simplex tree under landmark
insertions. Adding new vertices in witness complexes is used
in~\cite{DBLP:journals/dcg/BoissonnatGO09} for manifold reconstruction.
%
Given the set of landmarks $L$, the set of witnesses $W$ and the $k$-skeleton of the
witness complex $\wit(W,L)$ represented as a simplex tree, we take a new landmark point
$x$ and we update the simplex tree so as to construct the simplex tree
associated to $\wit(W,L\cup \{x\})$. We assign to $x$ the biggest label $|L|+1$. We suppose to
have at our disposal an oracle that can compute the subset $W^x
\subseteq W$ of the witnesses that
admit $x$ as one of their $k+1$ nearest neighbors. 
Computing $W^x$ is known as the \emph{reverse nearest neighbor} search
problem, which
has been intensively studied in the past few years~\cite{DBLP:conf/sigmod/AchtertBKKPR06}.
Let $w$ be a witness in $W^x$ and suppose $x$ is its $(i+1)^{\text{th}}$
nearest neighbor in $L\cup \{x\}$, with $0\leq i \leq k$. Let
$\sigma_j \subseteq L$
be the $j$-dimensional simplex witnessed by $w$ in $L$ and let
$\widetilde{\sigma_j} \subseteq L\cup \{x\}$ be the $j$-dimensional simplex
witnessed by $w$ in $L\cup \{x\}$. Consequently, $\sigma_j = \widetilde{\sigma_j}$ for $j < i$
and $\sigma_j \neq \widetilde{\sigma_j}$ for $j \geq i$.
We equip each node $N$ of the simplex tree with a \emph{counter of witnesses} which maintains
the number of witnesses that witness the simplex represented by $N$. As
for the witness complex construction, we consider all
nodes representing simplices witnessed by elements of $W^x$,
proceeding by increasing dimensions. 
For a witness $w\in W^x$ and a dimension  $j \geq
i$, we decrement the witness counter of $\sigma_j$ and insert
$\widetilde{\sigma_j}$ if and only if  its facets are in the simplex tree. We
remark that $[\widetilde{\sigma_j}] = [\sigma_{j-1}]\centerdot [x]$
because $x$ has the biggest label of all landmarks. We can thus access in time
$O(D_{\text{m}})$ the position of the word  $[\widetilde{\sigma_j}]$ since we
have accessed the node representing $[\sigma_{j-1}]$ in the previous iteration of
the algorithm.

If the witness counter of a node is turned down to $0$, the simplex
$\sigma$ it represents is not witnessed anymore, and is consequently
not part of $\wit(W,L\cup \{x\})$. We remove the nodes representing
$\sigma$ and its cofaces from the simplex tree, using {\sc Locate-cofaces}.

\paragraph{Complexity.}

The update procedure is  a ``local''  variant of the witness complex construction,
where, by ``local'',  we mean that we reconstruct only the star of vertex
$x$. Let $C_x$ denote  the number of cofaces of $x$ in $\wit(W,L\cup
\{x\})$ (or equivalently the size of its star).  The same analysis as above shows that
updating the simplicial complex takes time $O( (|W^x| +
C_x)k^2D_{\text{m}})$, plus one call to the oracle to compute $W^x$.

\paragraph{Relaxed Witness Complex.}
Given a relaxation parameter $\rho\geq 0$ we define the \emph{relaxed
  witness complex}~\cite{Alexa_topologicalestimation}:

\begin{definition}
A witness $w\in W$ \emph{$\rho$-witnesses} a simplex $\sigma \subseteq L$ iff:
$$\forall x \in \sigma \mbox{ and } \forall y \in L\setminus \sigma
\mbox{ we have } d(w,x) \leq d(w,y)+\rho$$
\end{definition}

The \emph{relaxed witness complex} \rwit$(W,L)$ with parameter $\rho$ is the maximal simplicial complex,
  with vertices in $L$, whose faces admit a $\rho$-witness in $W$. For
  $\rho=0$, the relaxed witness complex is the standard witness
  complex. The parameter $\rho$ defines a filtration on the witness
  complex, which has been used in topological data analysis.

We resort to the same incremental algorithm as above. At each step
$j$, we insert,
for each witness $w$, the $j$-dimensional
simplices which are $\rho$-witnessed by $w$. 
Differently from the standard witness complex, there may be more than
one $j$-simplex that is witnessed by a
given witness $w\in W$. Consequently, we do not
maintain a pointer from each active witness to the last inserted simplex
it witnesses. We 
use simple top-down insertions from the root of the simplex tree.

Given a witness $w$ and a dimension $j$, we generate all the $j$-dimensional simplices which are
$\rho$-witnessed by $w$. For the ease of exposition, we
suppose we are given the sorted list of
nearest neighbors of $w$ in $L$, noted $\{z_0 \cdots z_{|L|-1}\}$, and 
their distance to $w$,
 noted $m_i = \text{d}(w,z_i)$, with
$m_0\leq \cdots \leq m_{|L|-1}$, breaking ties arbitrarily. Note that if one
wants to construct only the $k$-skeleton of the complex, it is
sufficient to know the list of
neighbors of $w$ that are at distance at most $m_k +\rho$
from $w$. We preprocess this list of neighbors for all witnesses. 
For $i \in \{0, \cdots, |L|-1\}$, we define the set $A_i$ of
landmarks $z$ such that $ m_i \leq d(w,z) \leq m_i+\rho$. For $i \leq
j+1$, $w$ $\rho$-witnesses all the $j$-simplices that contain
$\{z_0, \cdots , z_{i-1}\}$ and a $(j+1-i)$-subset of $A_i$, provided
$|A_i|\geq j+1-i$. We see that all $j$-simplices that are
$\rho$-witnessed by $w$ are obtained this way, and exactly once, when
$i$ ranges from $0$ to $j+1$.

\begin{figure}
\begin{center}
\includegraphics[width=10cm]{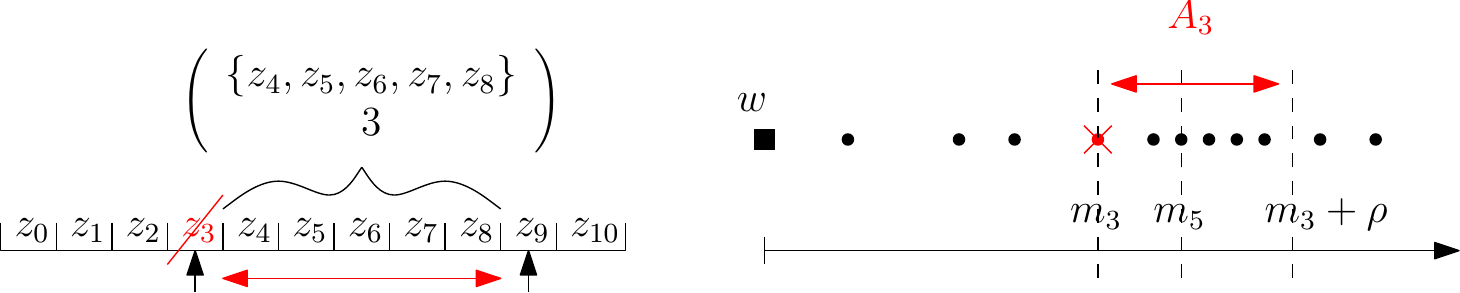}
\end{center}
\caption{Computation of the $\rho$-witnessed simplices $\sigma$ of
  dimension $5$. If $z_3$ is the first neighbor of $w$ not in
  $\sigma$, then $\sigma$ contains $\{z_0,z_1,z_2\}$ and any $3$-uplet
of $A_3 = \{z_4,\cdots,z_8\}$.}
\label{fig:relax-algo}
\end{figure}

For all $i \in \{0, \cdots , j+1\}$, we compute $A_i$ and generate
all the simplices which contain $\{z_0, \cdots , z_{i-1}\}$ and a 
subset of $A_i$ of size $(j+1-i)$. In order to easily update $A_i$ when $i$ is
incremented, we maintain two pointers to the list of neighbors, one
to $z_i$  and the other to the end of $A_i$. We check in constant time if
$A_i$ contains more than $j+1-i$ vertices, and compute all the
subsets of $A_i$ of cardinality $j+1-i$ accordingly. See Figure~\ref{fig:relax-algo}.

\paragraph{Complexity.} 
Let $R_j$ be the number of $j$-simplices $\rho$-witnessed by $w$.
Generating all those simplices takes $O(j + R_j)$ time. Indeed, for
all $i$ from $0$ to $j+1$, we construct $A_i$ and check whether $A_i$
contains more than $j+1-i$ elements. This is done by a simple
traversal of  the list of neighbors of $w$, which takes $O(j)$ time. Then, when
$A_i$ contains more than $j+1-i$ elements, we generate all subsets of
$A_i$ of size $j+1-i$ in time $O( \binom{|A_i|}{j+1-i} )$. As each
such subset leads to a $\rho$-witnessed simplex, the total cost
for generating all those simplices is $O(R_j)$.

We can deduce the complexity of the construction of the
relaxed witness complex. Let $\mathcal{R} = \displaystyle \sum_{w\in
  W} \sum_{j=0\cdots k} R_j$ be the number of $\rho$-witnessed
simplices we try to insert.  The construction of the relaxed witness complex takes
$O(\mathcal{R} k^2D_{\text{m}})$ operations. This bound is quite pessimistic
and, in practice, we observed that the construction time
is sensitive to the size of the output complex.
Observe that the quantity analogous to $\mathcal{R}$ in the case of
the standard witness complex was $k|W|$ and that the complexity was better
due to our use of the notion of active witnesses.

\section{Experiments}
\label{sec:experiments}

\begin{figure}
\begin{center}
\setlength{\tabcolsep}{3.72pt}
\begin{tabular}{|l | r  r  r  r  r  r  r  r  r  r|}
\hline
Data &  $|\mathcal{P}|$& $D$& $d$& $r$& $T_{\mbox{g}}$& $|E|$& $T_{\mbox{Rips}}$& $|\mathcal{K}|$& $T_{\mbox{tot}}$& $T_{\mbox{tot}}/|\mathcal{K}|$\\
\hline
{\bf Bud} & 49,990& 3& 2& 0.11& 1.5& 1,275,930& 104.5& 354,695,000& 104.6& $3.0 \cdot 10^{-7}$\\
{\bf Bro} & 15,000& 25& ?& 0.019& 0.6& 3083& 36.5& 116,743,000& 37.1& $3.2 \cdot 10^{-7}$\\
{\bf Cy8}& 6,040& 24& 2& 0.4& 0.11& 76,657& 4.5& 13,379,500& 4.61& $3.4 \cdot 10^{-7}$\\
{\bf Kl}   & 90,000& 5& 2& 0.075& 0.46& 1,120,000 & 68.1&  233,557,000& 68.5& $2.9 \cdot 10^{-7}$\\
{\bf S4} & 50,000& 5& 4& 0.28& 2.2&
1,422,490& 95.1& 275,126,000&  97.3 & $3.6 \cdot 10^{-7}$\\
\hline
\end{tabular}
\setlength{\tabcolsep}{3.6pt}
\begin{tabular}{|l|  r  r  r  r  r c r  r  r  r  r|}
\hline
Data &$|L|$ & $|W|$ & $D$ & $d$ & $\rho$ && $T_{\mbox{nn}}$ & $T_{\mbox{\rwit}}$ & $|\mathcal{K}|$ & $T_{\mbox{tot}}$ & $T_{\mbox{tot}}/|\mathcal{K}|$\\
\hline
{\bf Bud}
&10,000 & 49,990 & 3 & 2 & 0.12 && 1. &  729.6 & 125,669,000 & 730.6 & $0.58
\cdot 10^{-5}$\\
{\bf Bro}
&3,000 & 15,000 & 25 & ? & 0.01 && 9.9 & 107.6 & 2,589,860 & 117.5 & $4.5 \cdot 10^{-5}$\\
{\bf Cy8}&
800 & 6,040  &  24  & 2  & 0.23 &&  0.38
& 161 &  997,344  &  161.2  &  $16 \cdot 10^{-5}$ \\
{\bf Kl}
&10,000 & 90,000 & 5 & 2 & 0.11 && 2.2 & 572 & 109,094,000 & 574.2 & $0.53\cdot 10^{-5}$\\
{\bf S4} &  50,000 & 200,000    &  5  & 4   &  0.06  && 25.1   & 296.7  &  163,455,000  &  321.8   & $0.20 \cdot 10^{-5}$\\
\hline
\end{tabular}
\end{center}
\caption{Data, timings (in s.) and statistics for the construction of
 Rips complexes (TOP)  and relaxed witness complexes  (BOTTOM). All
 complexes are constructed up to embedding dimension.}
\label{fig:table_rips_wc}
\end{figure}
In this section, we report on the performance of our algorithms on both
real and synthetic data, and compare them to existing software. More
specifically, we benchmark the construction of Rips complexes, witness complexes and relaxed
witness complexes. Our implementations are in {\tt C++}. We use the {\sc ANN}
library~\cite{ann_mount} to compute the $1$-skeleton graph of the Rips
complex, and to compute the lists of nearest neighbors of the
witnesses for the witness complexes. All
timings are measured on a Linux machine with $3.00$ GHz processor and $32$
GB RAM. For its efficiency and flexibility, we use the {\tt map} container
of the {\tt Standard Template Library}~\cite{SGIguide} for storing
sets of sibling nodes, except for the top nodes which are stored in an
array. 

We use a variety of both real and synthetic datasets. {\bf Bud} is a
set of points sampled from the surface of the {\it Stanford Buddha}~\cite{buddha_stanford_scan} in
$\R^3$. {\bf Bro} is a set of $5\times 5$ {\it high-contrast patches}
derived from natural images, interpreted as vectors in $\R^{25}$, from the Brown database (with parameter $k=300$ and cut
$30\%$)~\cite{DBLP:journals/ijcv/CarlssonISZ08,DBLP:journals/ijcv/LeePM03}. {\bf Cy8} is a set of
points in $\R^{24}$, sampled from the space of conformations of the
cyclo-octane molecule~\cite{martin2010top}, which is the union of two
intersecting surfaces. {\bf Kl} is a set of points sampled from the
surface of the figure eight Klein Bottle embedded in $\R^5$. Finally
{\bf S4} is a set of points uniformly distributed on the unit $4$-sphere in
$\R^5$. Datasets are listed in Figure~\ref{fig:table_rips_wc} with
details on the sets of points $\mathcal{P}$ or landmarks $L$ and
witnesses $W$,
their size $|\mathcal{P}|$, $|L|$ and $|W|$, the ambient dimension $D$, the intrinsic
dimension $d$ of the object the sample points belong to (if known),
the parameter $r$ or $\rho$, the dimension $k$ up to which we construct
the complexes, the time $T_{\mbox{g}}$  to construct the Rips graph or
the time $T_{\mbox{nn}}$ to compute the lists of nearest neighbors of
the witnesses, the number of edges $|E|$, the time for the
construction of 
the Rips
complex $T_{\mbox{Rips}}$ or for the construction of the witness complex $T_{\mbox{\rwit}}$, the size of the complex
$|\mathcal{K}|$, and the total construction time $T_{\mbox{tot}}$ and
average construction time per face $T_{\mbox{tot}} / |\mathcal{K}|$.

We test our algorithms on these datasets, and
compare their performance with two existing softwares that are state-of-the-art. 
We compare our implementation to the
{\sc JPlex}~\cite{jplex_cite} library and the {\sc Dionysus}~\cite{dionysus_morozov}
library. The first is a Java package which can be used with
{\tt Matlab} and provides an implementation of the construction of
Rips complexes and witness complexes. The
second is implemented in {\tt C++} and provides an implementation of
the construction of Rips complexes. Both libraries are
widely used to construct simplicial complexes and to compute their
persistent homology.
We also provide an experimental analysis of the memory
performance of our data structure compared to other
representations. Unless mentioned otherwise, all simplicial complexes
are computed up to the embedding dimension.

All timings are
averaged over $10$ independent runs. Timings are provided by the {\tt clock} function from the {\tt
  Standard C Library}, and zero means that the measured time is below
the resolution of the {\tt clock} function. Experiments are stopped
after one hour of computation, and data missing on plots means that
the computation ran above this time limit.


For readability, we do not report on
the performance of each algorithm on each dataset in this section, but the 
results presented are a faithful sample of what we have
observed on other datasets. A complete set of experiments is reported
in appendix~\ref{sec:additional_expe}.


As illustrated in Figure~\ref{fig:table_rips_wc}, we are able to
construct and represent both Rips and relaxed witness complexes of up to several hundred million
faces in high dimensions, on all datasets.


\paragraph{Data structure in {\sc JPlex} and {\sc Dionysus}:} 
Both {\sc JPlex} and {\sc Dionysus} represent the combinatorial
structure of a simplicial complex by its \emph{Hasse diagram}.
The \emph{Hasse diagram} of a
simplicial complex $\K$ is the graph whose nodes are in
bijection with the 
simplices (of all dimensions) of the simplicial complex and where an edge
links two nodes representing two simplices $\tau$ and $\sigma$ iff 
$\tau \subseteq \sigma$ and $\Dim{\sigma} = \Dim{\tau}+1$.

 {\sc JPlex} and {\sc Dionysus} are
 libraries dedicated to topological data analysis, where only the
 construction of simplicial complexes and the computation of the
 facets of a simplex are necessary.

For a simplicial complex $\K$ of dimension $k$ and a simplex $\sigma
\in \K$
of dimension $j$, the Hasse diagram has
size $\Theta(k|\K|)$ and allows to compute {\sc Locate-facets}$(\sigma)$
in time $O(j)$, whereas the simplex tree has size $\Theta(|\K|)$ and
allows to compute {\sc Locate-facets}$(\sigma)$ in
time $O(j^2D_{\text{m}})$.


\subsection{Memory Performance of the Simplex Tree}

 \begin{figure}

\setlength{\tabcolsep}{13pt}
 \hspace{-0.5cm}\begin{tabular}{c c}
    \begin{minipage}[b]{0.45\linewidth}
      \centering
      \includegraphics{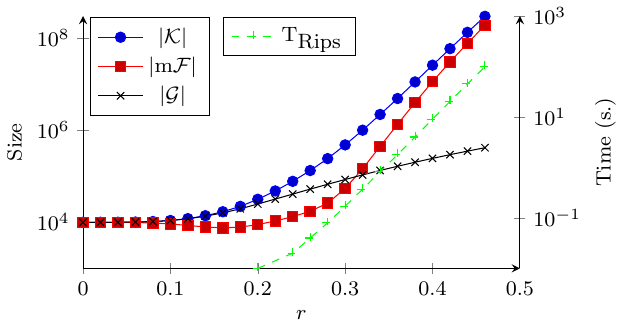}
    \end{minipage}
    &  
    \begin{minipage}[b]{0.45\linewidth}
      \centering
      \includegraphics{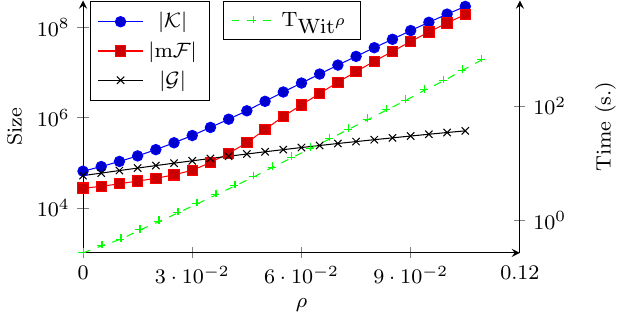}
    \end{minipage}\\
  \end{tabular}

\caption{Statistics and timings for the Rips complex (Left)
  and the relaxed witness complex (Right) on {\bf S4}.} 
\label{fig:size_time}
\end{figure}

In order to represent the combinatorial structure of an arbitrary
simplicial complex, one needs to mark all maximal faces. Indeed,
except in some special cases (like in flag complexes where all faces
are determined by the $1$-skeleton of the complex), one cannot infer
the existence of a simplex in a simplicial complex $\K$ from the
existence of its faces in $\K$.
Moreover, the number of
maximal simplices of a $k$-dimensional simplicial complex is at least
$|V| /(k+1)$. In the case, considered in this paper, where
the vertices are identified by their labels, a  
minimal representation of the maximal simplices would then require at
least $\Omega (\log |V|)$ bits per
maximal face, for fixed $k$.
The simplex tree 
uses $O(\log |V|)$ memory bits per face {\em of any dimension}. The following
experiment compares the memory performance of the simplex tree with
the minimal representation described above, and with the
representation of the $1$-skeleton.

Figure~\ref{fig:size_time} shows results for both Rips 
and relaxed witness complexes associated to $10,000$ points from
{\bf S4}
and various values of, respectively, the distance threshold $r$ and the
relaxation parameter $\rho$. 
The figure plots the total number of faces $|\mathcal{K}|$, the number of
maximal faces $|\mbox{m}\mathcal{F}|$,
the size of the $1$-skeleton $|\mathcal{G}|$ and the construction times
$T_{\mbox{Rips}}$ and $T_{\mbox{\rwit}}$.

As expected, the $1$-skeleton is significantly smaller than the two
other representations. However, as explained earlier, a representation
of the graph of the simplicial complex is only well suited for
flag complexes. 

As shown on the figure, the total number of faces and the number of maximal faces remain
close along the experiment. Interestingly, we catch the topology of {\bf S4} when
 $r\approx 0.4$ for the Rips complex and $\rho \approx 0.08$ for the relaxed witness
  complex. For these ``good'' values of the parameters, the total
  number of faces is not much bigger than the number of maximal
faces. Specifically, the total number of faces of the Rips complex is less than $2.3$
times bigger than the number of maximal faces, and the ratio is less than $2$
for the relaxed witness complex.

\subsection{Construction of Rips Complexes}
\begin{figure}

\setlength{\tabcolsep}{13pt}
 \hspace{-0.5cm}\begin{tabular}{c c}
    \begin{minipage}[b]{0.45\linewidth}
      \centering
      \includegraphics{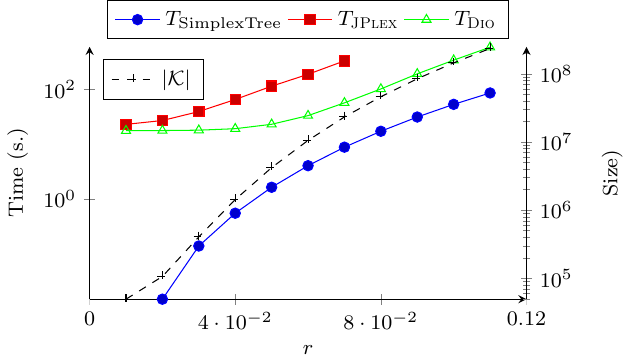}
    \end{minipage}
    &  
    \begin{minipage}[b]{0.45\linewidth}
      \centering
      \includegraphics{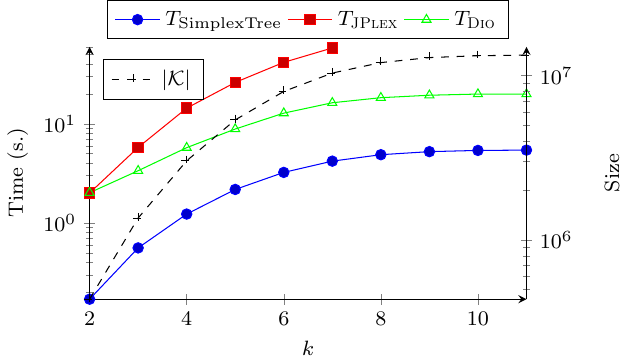}
    \end{minipage}\\
  \end{tabular}

\caption{Statistics and timings for the construction of the Rips
    complex on (Left) {\bf Bud} and (Right) {\bf Cy8}.} 
\label{fig:plots_time_rips}
\end{figure}


We test our algorithm for
the construction of Rips complexes.
In Figure~\ref{fig:plots_time_rips} we compare the performance of our
algorithm with {\sc JPlex} and with {\sc Dionysus} along two directions.

In the first experiment, we build the Rips complex on $49,000$ points
from the dataset {\bf Bud}. Our construction is
at least $36$ times faster than {\sc JPlex} along the experiment, and
several hundred times faster for small values of the parameter $r$. Moreover, {\sc
  JPlex} is not able to handle the full dataset {\bf Bud} nor big simplicial complexes due to memory
allocation issues, whereas our method has no such problems. In our experiments, {\sc
  JPlex} is not able to compute complexes of more than $23$ million faces ($r=0.07$)
while the simplex tree construction runs successfully until
$r=0.11$, resulting in a complex of $237$ million faces.
Our construction is at least $7$ times faster than {\sc Dionysus}
along the experiment, and several hundred times faster for small
values of the parameter $r$.

In the second experiment, we construct the Rips complex on the $6040$
points from {\bf Cy8},
with threshold $r=0.4$, for different
dimensions $k$. Again, our method outperforms {\sc JPlex}, by a factor
$11$ to $14$. {\sc JPlex} cannot compute complexes of dimension higher than
$7$ because it is limited by design to simplicial complexes of dimension smaller than $7$.
Our construction is $4$ to $12$ times faster than {\sc Dionysus}.

The simplex tree and the expansion algorithm we have described are
output sensitive. 
As shown by our experiments, the construction time using a simplex tree depends linearly on the size of the output complex. Indeed, when the Rips graphs are dense enough
so that the time for the expansion dominates the full
construction, we observe that the average construction time per face is
constant and equal to
$3.7\times 10^{-7}$ seconds for the first experiment, and $4.1\times
10^{-7}$ seconds for the second experiment (with standard errors
$0.20\%$ and $0.14\%$ respectively).

\subsection{Construction of Witness Complexes}
%
\begin{figure}

\setlength{\tabcolsep}{13pt}
 \hspace{-0.5cm}\begin{tabular}{c c}
    \begin{minipage}[b]{0.45\linewidth}
      \centering
      \includegraphics{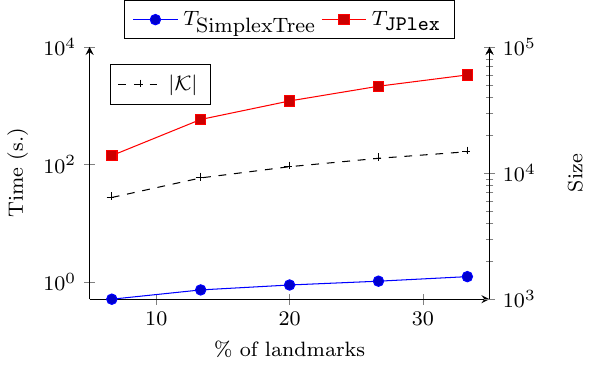}
    \end{minipage}
    &  
    \begin{minipage}[b]{0.45\linewidth}
      \centering
      \includegraphics{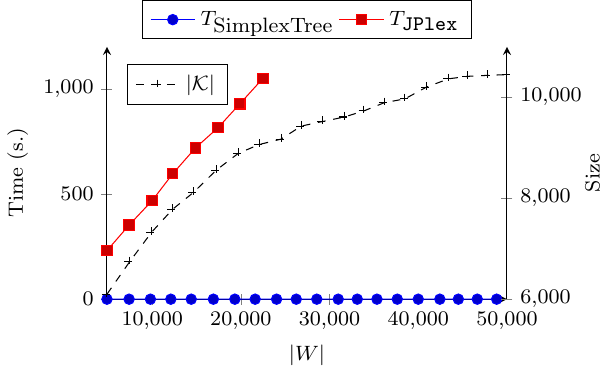}
    \end{minipage}\\
 
\begin{minipage}[b]{0.45\linewidth}
      \centering
      \includegraphics{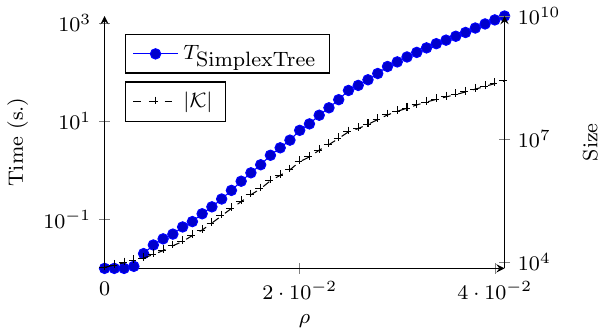}
    \end{minipage}
    &  
    \begin{minipage}[b]{0.45\linewidth}
      \centering
      \includegraphics{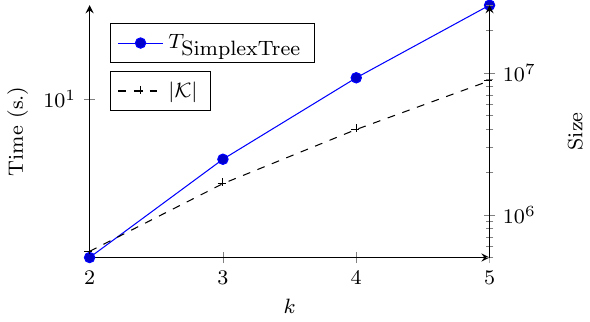}
    \end{minipage}\\

  \end{tabular}

\caption{Statistics and timings for the construction of: (TOP) the
  witness complex and (BOTTOM) the relaxed witness complex, on
  datasets (Left) {\bf Bro} and (Right) {\bf Kl}.}
\label{fig:non-relaxed-wc--new}
\end{figure}

We test our algorithms for the construction of witness complexes and
relaxed witness complexes. 

Figure~\ref{fig:non-relaxed-wc--new} (top) shows the results of two
experiments for the full construction of witness complexes. The first
one compares the performance of the simplex tree algorithm and
of {\sc JPlex} on the dataset {\bf Bro} consisting of $15,000$
points in dimension $\R^{25}$. Subsets of different size of landmarks are selected at random
among the sample points. 
Our algorithm is from several hundred to several thousand times faster
than {\sc JPlex} (from small to big subsets of landmarks). Moreover, the simplex
tree algorithm for the construction of the witness complex represent
less than $1\%$ of the total time spent, when more than $99\%$ of the
total time is spent computing the
nearest neighbors of the witnesses.

In the second experiment, we construct the witness complex on $2,500$
landmarks from {\bf Kl}, and sets of witnesses of different size.
The simplex tree algorithm outperforms {\sc JPlex}, being tens of
thousands times faster. {\tt JPlex} runs above the one hour time limit
when the simplex tree algorithm stays under
$0.1$ second all along the experiment. Moreover, the simplex
tree algorithm spends only about $10\%$ of the time constructing the
witness complex, and $90\%$ computing the
nearest neighbors of the witnesses.

Finally we test the full construction of the relaxed witness complex.
{\sc JPlex} does not provide an implementation of the relaxed
witness complex as defined in this paper; consequently, we were not
able to compare the algorithms on the construction of the relaxed witness complex.
We test our algorithms along two directions, as illustrated in Figure~\ref{fig:non-relaxed-wc--new} (bottom).
In the first experiment, we compute the $5$-skeleton of the relaxed
witness complex on {\bf Bro}, with $15,000$ witnesses and $1,000$
landmarks selected randomly, for different values of the parameter
$\rho$. 
In the second experiment, we construct the $k$-skeleton of the relaxed witness complex on
{\bf Kl} with $10,000$ landmarks, $100,000$ witnesses and fixed
parameter $\rho=0.07$, for various
$k$. We are able to construct and store complexes of up to $260$ million faces.
In both cases the construction time is linear in the size of the
output complex, with a contruction time per face equal to $4.9 \times
10^{-6}$ seconds in the first experiment, and $4.0 \times 10^{-6}$
seconds in the second experiment (with standard errors $1.6\%$ and
$6.3\%$ respectively).

\section*{Conclusion}
\label{sec:conclusion}

 We believe that the simplex tree is the first scalable and truly practical data structure
 to represent general simplicial complexes. The simplex tree is very
 flexible, can represent any kind of simplicial complexes and allow
 efficient implementations of all basic operations on simplicial
 complexes. 
Futhermore, since the simplex tree stores all simplices of the
simplicial complex, it has been successfully applied to represent
filtrations and to compute persistent homology~\cite{boissonnat:hal-00761468}.
We plan to make our code publicly available and to use it for practical
 applications in data analysis and manifold learning.
Further developments also include more compact storage
using succinct representations of
trees~\cite{DBLP:conf/focs/Jacobson89}.


\begin{acknowledgements}
 The authors thanks A.Ghosh, S. Hornus, D. Morozov and P.  Skraba for
 discussions that led to the idea of representing simplicial complexes
 by tries. They especially thank S. Hornus for sharing his notes with
 us
. They also thank S. Martin and
 V. Coutsias for providing the cyclo-octane data set. 
This research has been partially supported by the 7th Framework Programme for Research of the European Commission, under FET-Open grant number 255827 (CGL Computational Geometry Learning).
\end{acknowledgements}

\bibliographystyle{plain}
\bibliography{bibliography.bib}

 \newpage

 \appendix
 \section{Additional Experiments}
 \label{sec:additional_expe}

In this section we provide more experiments on the running time of the
algorithms for constructing Rips complexes and relaxed witness
complexes on all datasets. The datasets used are described in Figure~\ref{fig:table_rips_wc}.

\begin{figure}[h]%
  \centering
  \setlength{\tabcolsep}{5pt}
    \begin{tabular}{l  r | r r r r r r r}
      {\bf Bud:} & $r$ & $0.08$ & $0.085$ & $0.090$ & $0.095$ & $0.100$ & $0.105$ & $0.110$ \\ \cline{2-9}
      & $T_{\text{Rips}}$ & $19.4$ &  $26.5$ & $35.8$ & $46.7$ &
      $60.5$ & $77.7$ & $98.7$ \\

\vspace{0.3cm}

      & $|\K|$ & $69 \cdot 10^{6}$ & $94 \cdot 10^{6}$ & $127 \cdot 10^{6}$ & $167 \cdot 10^{6}$ & $217 \cdot 10^{6}$ & $280 \cdot 10^{6}$ & $355 \cdot 10^{6}$ \\

      {\bf Bro:} & $r$ & $0.184$ & $0.186$ & $0.188$ & $0.190$ & $0.192$ & $0.194$ & $0.196$\\ \cline{2-9}
      & $T_{\text{Rips}}$ & $15.3$ &  $18.1$ & $28.2$ & $34.5$ & $40.8$ & $56.2$ & $81.1$ \\
\vspace{0.3cm}

      & $|\K|$ & $52 \cdot 10^{6}$ &  $61 \cdot 10^{6}$ & $95 \cdot 10^{6}$ & $117 \cdot 10^{6}$ & $138 \cdot 10^{6}$ & $190 \cdot 10^{6}$ & $275 \cdot 10^{6}$ \\

      {\bf Cy8:} & $r$ & $0.406$ & $0.415$ & $0.424$ & $0.433$ & $0.442$ & $0.451$ & $0.460$\\ \cline{2-9}
      & $T_{\text{Rips}}$ & $5.7$ & $8.7$ & $13.6$ & $21.4$ & $34.5$ & $57.3$ & $96.6$\\
\vspace{0.3cm}

      & $|\K|$ & $17 \cdot 10^{6}$ & $27 \cdot 10^{6}$ & $42 \cdot 10^{6}$ & $67 \cdot 10^{6}$ & $108 \cdot 10^{6}$ & $180 \cdot 10^{6}$ & $305 \cdot 10^{6}$\\

      {\bf Kl:} & $r$ & $0.059$ & $0.062$ & $0.065$ & $0.068$ & $0.071$ & $0.074$ & $0.077$\\ \cline{2-9}
      & $T_{\text{Rips}}$ & $7.0$ & $11.1$ & $17.8$ & $26.3$ & $38.4$ & $58.3$ & $87.3$\\
\vspace{0.3cm}

      & $|\K|$ & $24 \cdot 10^{6}$ & $38 \cdot 10^{6}$ & $61 \cdot 10^{6}$ & $90 \cdot 10^{6}$ & $133 \cdot 10^{6}$ & $204 \cdot 10^{6}$ & $305 \cdot 10^{6}$\\
 
      {\bf S4:} & $r$ & $0.22$ & $0.23$ & $0.24$ & $0.25$ & $0.26$ & $0.27$ & $0.28$\\ \cline{2-9}
      & $T_{\text{Rips}}$ & $2.7$ & $4.7$ & $8.5$ & $15.4$ & $28.0$ & $50.9$ & $93.7$\\
\vspace{0.3cm}

      & $|\K|$ & $7 \cdot 10^{6}$ & $13 \cdot 10^{6}$ & $23 \cdot 10^{6}$ & $43 \cdot 10^{6}$ & $79 \cdot 10^{6}$ & $146 \cdot 10^{6}$ & $271 \cdot 10^{6}$\\
    \end{tabular}

\caption{Timings $T_{\text{Rips}}$ for the construction of the Rips
  complex on the data
  sets and size of the simplicial complexes $|\K|$, for different
  values of the parameter $r$. On all these
experiments, the time complexity is linear in the number of
faces. Specifically, the timing per simplex ranges between $2.79 \cdot
10^{-7}$ and $3.47\cdot 10^{-7}$ seconds per simplex depending on the
dataset, with standard
error at most $0.40\%$.}
\label{add_expe:rips}
\end{figure}
\begin{figure}
  \centering
  \setlength{\tabcolsep}{5pt}
    \begin{tabular}{l  r | r r r r r r r}
     {\bf Bud:} & $\rho$ & $0.06$ & $0.07$ & $0.08$ & $0.09$ & $0.10$ & $0.11$ & $0.12$\\ \cline{2-9}
      & $T_{\mbox{\rwit}}$ & $18.3$ & $36.9$ & $71.1$ & $135.8$ & $249.1$ & $440.2$ & $758.6$\\
\vspace{0.3cm}

      & $|\K|$ & $7.8 \cdot 10^{6}$ & $14 \cdot 10^{6}$ & $23 \cdot 10^{6}$ & $38 \cdot 10^{6}$ & $58 \cdot 10^{6}$ & $88 \cdot 10^{6}$ & $130 \cdot 10^{6}$\\

     {\bf Bro:} & $\rho$ & $0.0075$ & $0.0080$ & $0.0085$ & $0.0090$ & $0.0095$ & $0.0100$ & $0.0105$\\ \cline{2-9}
      & $T_{\mbox{\rwit}}$ & $4.0$ & $6.1$ & $10.7$ & $16.5$ & $39.3$ & $123.2$ & $530.9$\\
\vspace{0.3cm}

      & $|\K|$ & $1.2 \cdot 10^{6}$ & $1.5 \cdot 10^{6}$ & $1.9 \cdot 10^{6}$ & $2.2 \cdot 10^{6}$ & $3.1 \cdot 10^{6}$ & $4.6 \cdot 10^{6}$ & $7.0 \cdot 10^{6}$\\

     {\bf Cy8:} & $\rho$ & $0.194$ & $0.200$ & $0.206$ & $0.212$ & $0.218$ & $0.224$ & $0.230$\\ \cline{2-9}
      & $T_{\mbox{\rwit}}$ & $18.7$ & $33.1$ & $130.2$ & $273.0$ & $512.9$ & $37.2$ & $1411.2$\\
\vspace{0.3cm}

      & $|\K|$ & $0.45 \cdot 10^{6}$ & $0.66 \cdot 10^{6}$ & $0.82 \cdot 10^{6}$ & $1.1 \cdot 10^{6}$ & $1.7 \cdot 10^{6}$ & $2.3 \cdot 10^{6}$ & $3.6 \cdot 10^{6}$\\

     {\bf Kl:} & $\rho$ & $0.05$ & $0.06$ & $0.07$ & $0.08$ & $0.09$ & $0.10$ & $0.11$\\ \cline{2-9}
      & $T_{\mbox{\rwit}}$ & $3.2$ & $9.7$ & $24.6$ & $55.3$ & $118.0$ & $261.1$ & $584.5$\\
\vspace{0.3cm}

      & $|\K|$ & $0.78 \cdot 10^{6}$ & $2.2 \cdot 10^{6}$ & $5.2 \cdot 10^{6}$ & $11 \cdot 10^{6}$ & $23 \cdot 10^{6}$ & $49 \cdot 10^{6}$ & $109 \cdot 10^{6}$\\

     {\bf S4:} & $\rho$ & $0.03$ & $0.035$ & $0.040$ & $0.045$ & $0.050$ & $0.055$ & $0.060$\\ \cline{2-9}
      & $T_{\mbox{\rwit}}$ & $7.6$ & $14.1$ & $26.4$ & $48.9$ & $89.2$ & $164.6$ & $297.3$\\
\vspace{0.3cm}

      & $|\K|$ & $2.8 \cdot 10^{6}$ & $5.3 \cdot 10^{6}$ & $11 \cdot 10^{6}$ & $22 \cdot 10^{6}$ & $43 \cdot 10^{6}$ & $85 \cdot 10^{6}$ & $161 \cdot 10^{6}$\\
\end{tabular}

\caption{Timings $T_{\text{\rwit}}$ for the construction of the
  relaxed witness
  complex on the data
  sets and size of the simplicial complexes $|\K|$, for different
  values of the parameter $\rho$. The
timings per simplex vary, as the complexity of the construction
algorithm depends also on the number of witnesses. It however ranges
between $\approx 10^{-6}$ and $\approx 10^{-4}$ seconds per simplex
depending on the number of witnesses compared to the number of
simplices of the output.}
\label{add_expe:rwc}
\end{figure}

\end{document}